# MRI-Guided High Intensity Focused Ultrasound of Liver and Kidney

B Denis de Senneville[1], M Ries[1], LW Bartels[2], CTW Moonen[1,2]

[1]Laboratory for Molecular and Functional Imaging
CNRS/ University Segalen Bordeaux
146 rue Leo Saignat, Case 117
33076 Bordeaux, France

[2]Image Sciences Institute
Dept. of Radiology
University Medical Center Utrecht
The Netherlands

baudouin@imf.u-bordeaux2.fr
ries@imf.u-bordeaux2.fr
wilbert@isi.uu.nl
chrit@imf.u-bordeaux2.fr

**Acknowledgments:**
This authors acknowledge support from Agence National de Recherche (project MRgHIFU-ALKT), Fondation InNaBioSanté (project ULTRAFITT), Center for Translational Molecular Medicine (project VOLTA), Ligue Nationale Contre le Cancer, Conseil Régional d'Aquitaine and Philips Healthcare.

**Table of contents**

**Abstract**

High Intensity Focused Ultrasound (HIFU) can be used to achieve a local temperature increase deep inside the human body in a non-invasive way. MRI guidance of the procedure allows *in situ* target definition. In addition, MRI can be used to provide continuous temperature mapping during HIFU for spatial and temporal control of the heating procedure and prediction of the final lesion based on the received thermal dose. Temperature mapping of mobile organs as kidney and liver is challenging, as well as real-time processing methods for feedback control of the HIFU procedure. In this paper, recent technological advances are reviewed in MR temperature mapping of these organs, in motion compensation of the HIFU beam, in intercostal HIFU sonication, and in volumetric ablation and feedback control strategies. Recent pre-clinical studies have demonstrated the feasibility of each of these novel methods. The perspectives to translate those advances into the clinic are addressed. It can be concluded that MR guided HIFU for ablation in liver and kidney appears feasible but requires further work on integration of technologically advanced methods.

1. **Introduction**

Liver and kidney tumors represent a major health problem because few patients are eligible for curative treatment with surgery, and radio- and chemotherapy have shown limited success. Currently, radio-frequency ablation is the most used method for percutaneous treatment. The development of a completely non-invasive method based on MR guided high intensity focused ultrasound (HIFU) is of particular interest since the approach avoids some limitations associated with (mini-)invasive surgery.

Ultrasound allows the deposition of thermal and mechanical energy deep inside the human body in a non-invasive way. Since ultrasound can be focused within a region of about 1 mm diameter, focused ultrasound (FUS) opens a path towards new therapeutic strategies with improved reliability and reduced associated trauma. Current clinical application of FUS are in the field of direct tissue ablation by local temperature increase (hence the often used term High Intensity Focused Ultrasound, HIFU). The mechanical energy is transformed into thermal energy because of ultrasound-tissue interactions leading to local friction and relaxation.

The development of the non-invasive FUS heating technique and the excellent soft-tissue contrast and potential of MRI for on-line temperature mapping have provided the basis for the combination of the two technologies (Cline et al 1992) hereafter referred to as MRI-HIFU.

MRI-HIFU has been applied for ablation in immobilized organs (in particular, uterus, prostate, breast, brain). The procedures remain long (several hours) because the focal area of the ultrasonic waves has a small size (few $mm^3$). Thermal ablation using MRI-HIFU or ultrasound imaging has been evaluated in clinic for the treatment of various types of tumors, such as prostate (Gelet et al 1999), uterine fibroid (Stewart et al 2003, Zhang et al 2010, Tempany et al 2003, Stewart et al 2006) as well as breast (Zippel and Pappa 2005, Furusawa 2007), kidney cancer (Illing et al 2005, Hacker et al 2006) and liver cancer (Kennedy et al 2004), demonstrating that MR temperature mapping is indeed feasible for guiding HIFU therapy. The FDA approval was obtained in 2004 for treatment of uterine fibroids developed by GE/Insightec using a point-by-point approach. Philips has received European Community approval in 2009 for a volumetric HIFU approach for uterine fibroids. The HIFU technology is also under rapid development in the brain cancer research field, e.g. with the treatment of intracranial solid neoplasms trough the closed cranium (Ram et al 2006, Pernot et al 2003, Hynynen et al 2006).

The treatment of liver and kidney cancer with MRI-HIFU has so far been hampered by the complications arising from the physiological motion of both organs and their anatomical location: Directly located below the diaphragm, respiratory or cardiac-induced organ displacements and deformations will modify the local magnetic field experienced by the target organ. This generally complicates precise MR-thermometry with sufficient spatio-temporal resolution. Furthermore, continuous sonications with a non-invasive ablation system require a real-time motion compensation of the HIFU-beam. The focal point of the HIFU-system must be adaptively repositioned as the organ moves with respect to the external transducer in order to avoid damaging healthy tissue and to limit acoustic energy losses. In addition, ablations in the superior part of the abdomen are complicated by the partial obstruction of the HIFU beam path by the thoracic cage. This introduces both the risk of undesired tissue damage in the vicinity of bony and cartilaginous structures and degrades the focus quality and intensity. Finally, the treatment of large tumor volumes in highly perfused organs, and in particular the imperative to achieve a complete destruction of the tumor, require more efficient ablation control strategies compared to benign tumors, such as uterine fibroids.

Although only few selected cases of a successful treatment of liver or kidney cancer with MRI-HIFU have so far been demonstrated, many of these technical challenges which hamper widespread clinical adoption have meanwhile been addressed by methodological advances. This article aims to give an overview of the state of the field and an outlook to future developments.

2. **Motion compensation of the HIFU beam**

HIFU ablations in organs in the upper abdomen are challenging since the respiratory cycle causes a continuous displacement of the target area. Since both the liver and the kidney of a adult patient move under free-breathing conditions at rest with a period of 3-5 s and a motion amplitude of 10-20 mm, an uncompensated and continuous HIFU ablation will increase the risk of undesired tissue damage and reduce the treatment efficiency due to the spread of the acoustic energy along the target trajectory. Two principal approaches have been suggested to address this problem: Respiratory gated and continuous motion compensated sonication strategies.

2.1. Respiratory gated sonication strategies

As shown in Figure 1, the respiratory cycle of an adult patient leaves a temporal window of 1-2s within the respiratory cycle in which both liver and kidney remain stationary. Respiratory gated sonication strategies deposit the acoustic energy periodically with ablation bursts during this stationary interval (Okada et al 2006). The advantage of this approach is that no dynamic tracking of the exact organ position during inhalation and exhalation is required and that the HIFU system does not require real-time beam steering capabilities (Okada et al 2006).

On the other hand, respiratory gating significantly reduces the overall duty-cycle of the sonication process. This is particularly unfavorable for organs such as kidney and liver, since the high perfusion rate generally leads to a strong heat evacuation. In practice, this limits the realization of a sufficiently high temperature elevation to induce necrosis in larger volumes (see Figure 2) (Cornelis et al 2010). As a consequence, the small ablation volumes per burst are imperative to assure a complete destruction of tumor through overlap between the volumes and lead to long treatment times (Okada et al 2006). The use of very high acoustic pressures may lead to cavitation resulting in modification of focal point (location and shape) as well as in ultrasound energy absorption.

2.2. Continuous and motion compensated sonication strategies

Sustained sonication strategies allow depositing acoustic energy in the target area during the entire duration of respiratory cycle. As a consequence, compared to gated sonications they display a two to three-fold duty cycle improvement, which in turn shortens the total duration of the intervention (see Figure 1). However, this approach requires the focal point position to be continuously re-adjusted to the current target position in order to prevent both undesired tissue damage and energy spread. This is technically challenging, since the current organ position needs to be determined continuously with a high temporal and spatial resolution and the HIFU-beam position to be re-adjusted correspondingly with a sufficiently low latency. Recent studies have shown that dynamic beam steering with a temporal resolution of 10Hz and a spatial resolution below 2mm, in conjunction with a HIFU-beam re-adjustment latency of 100-150ms, generally allow to remove the effect of respiratory motion on the sonication (Ries et al 2010).

In order to achieve such a beam steering performance, two types of tracking approaches have been suggested: Indirect motion tracking and direct real-time motion tracking.

2.2.1. Beam steering based on indirect motion tracking

Since a direct observation of the 3D organ displacement with real-time MRI under the constraint of the required latency is technically challenging, initial approaches suggested exploiting the inherent periodicity of the respiratory induced motion pattern to separate the motion registration and the subsequent data analysis from the actual application of the correction during ablation process.

Denis de Senneville et al suggested (Denis de Senneville et al 2007) to pre-record a MR-image collection in a preparation scan and to subsequently calculate the target displacements off-line. During the HIFU ablation, the beam correction is applied in real-time from the "learned" motion model.

Since the preparation scan spans over several motion cycles in order to achieve sufficient sampling density, it is not necessary to employ ultra-fast MRI as long as intra-scan scan artifacts are avoided. This in turn, allows increasing both spatial resolution and volume coverage. Furthermore, since the preparation scan for the characterization of the motion is separated from MR-thermometry, this also opens the opportunity to employ different MR-sequences for each task: The preparation scan can be obtained with $T_1$ and $T_2$ weighted MR-sequences displaying excellent soft-tissue contrast and minimal image distortion (Barkhausen et al 2001) and the ablation process can be guided using gradient recalled MRI with long echo times for optimal MR-thermometry. Parallel to the acquisition of the MR-images, the signal of a real-time compatible external sensor, which characterizes the respiratory state, needs to be recorded. For this, different sensor types have been suggested in the past: Several groups (Moricawa et al 2002, Okada et al 2006) used a pneumatic sensor measuring the expansion of the thoracic cage for this purpose, while more recent approaches employed a subset of the MR-images (Denis de Senneville et al 2007), dedicated pencil-beam MR-navigator echoes (Nehrke et al 1999, Köhler et al 2011) or ultrasonic echoes (Günther and Feinberg et al 2004, Feinberg et al 2010) to observe the position of the diaphragm directly.

After completion of the pre-recording of the MR-images and the sensor data, the images need to be analyzed in order to relate the coordinate of each part of tissue in the image collection with the corresponding one in the reference image used for the interventional planning (Lourenço de Oliveira et al 2010). Several rapid image-processing techniques to estimate organ displacements from anatomical MR-images have been proposed in the past and an extensive review of the proposed methodology can be found elsewhere (Maintz and Viergever 1998). In the scope of MR-guided HIFU, initial approaches employed affine transformations for this purpose (Denis de Senneville et al 2004). Since affine image registration methods have a limited ability to address complex deformations, which are during the respiratory cycle predominantly present in the superior part of the liver, more recent approaches focused on methods which allow a motion estimation on a voxel-by-voxel basis such as optical flow based approaches (Denis de Senneville et al 2004, Denis de Senneville et al 2007, Roujol et al 2011) or pattern matching (Ross et al 2008). The result of this image registration is a motion vector field for each part of the respiratory cycle, which relates the position of each voxel in space to the reference position. In order to exploit this information in real-time for beam correction, it needs to be associated through a suitable model with the corresponding external sensor readings, which will provide during the HIFU ablation the required real-time information about the respiratory state.

Initial approaches stored the motion vector fields together with the external sensor readings in a look-up table for this purpose (Denis de Senneville et al 2007). During HIFU treatment, the current sensor reading is matched with the values in the look-up table and the corresponding displacement field was recovered. Since this approach requires large table sizes to achieve sufficient temporal sampling density to avoid aliasing artifacts and is intrinsically not able to correct for displacement amplitudes which have not been present during the preparation scan, several refinements have been proposed to address these shortcomings: Similar to existing approaches in the field of guided radiation therapy (Ernst et al 2007, Ramrath et al 2007), a parametric motion model based on harmonic functions was used for trajectory interpolation and extrapolation (Denis de Senneville et al 2007). Recently suggested methods employ a PCA-based motion descriptor to synthesize in real-time the complex organ deformation during the therapy (Maclair et al 2007, Denis de Senneville et al 2011). This approach allows in addition to maintain only the physiological components of the estimated motion in the parametric motion model, which permitted to reduce the noise on the estimated displacement.

Common to all these approaches is that during the subsequent HIFU ablation, the current state of the respiratory cycle is evaluated in real-time with help of the external sensor and the corresponding position of the target area is recovered through the parametric motion model and applied as a correction to the HIFU system. Since the method relies on the periodicity of the organ motion pattern which is induced by the respiratory cycle, it is particularly well suited for sedated or anesthetized patients with mechanically assisted respiration. The principal advantage of this approach is that the correction can be applied with a very high temporal resolution and a short tracking latency, which in turn reduces both undesired tissue damage and energy loss. Furthermore, since this approach restricts the required real-time processing to the ablation phase of the intervention, while calibration and modeling are carried out off-line, it is technically less demanding than direct tracking approaches with respect to motion estimation, MR-data acquisition and processing throughput. As a consequence this also potentially allows observing and characterizing the organ displacement completely in 3D and with high spatial resolution, which is hard to achieve in real-time.

On the other hand, indirect motion tracking does not observe the organ displacement directly, but relies instead on a calibration obtained prior to the intervention. Therefore, gradual changes in the motion pattern during the intervention, sensor drifts and spontaneous motion events require in practice a re-calibration periodically and thus limit the feasibility of the approach for extended durations of several minutes, in particularly when applied to patients under free-breathing conditions.

2.2.2. Beam steering based on direct real-time motion tracking

The performance of modern state-of-the-art MRI systems in conjunction with advanced image-processing techniques allows continuous real-time target identification and tracking with sub-scond temporal resolutions, even in the presence of complex motion patterns (Saborowski and Saaed 2007). Contrary to indirect motion tracking, this method performs the target localization directly on the continuously acquired MR-image data stream. This allows to detect spontaneous motion events immediately and to adapt to changes in the motion pattern dynamically without the requirement of a calibration prior or during the intervention. This renders the approach particularly suitable for the treatment of patients under free-breathing conditions and for sustained HIFU-ablations of long duration.

The high requirements with respect to real-time data acquisition and processing limited initial approaches to the observation of one-dimensional MRI information to locate the target: de Zwart et al (de Zwart et al 2001) suggested a MR-navigator-echo based motion tracking. The drawback of these initial direct motion tracking methods is their limitation to one or two dimensions and, due to the processing latency requirements, to affine registration algorithms only. Since affine image registration methods have a limited ability to address complex deformations, which are present in the superior part of the liver during the respiratory cycle, more recent approaches focused on methods which allow motion estimation based on a voxel-by-voxel basis, such as optical flow based approaches (Denis de Senneville et al 2007). However, the complexity of these algorithms generally leads to processing latencies which are incompatible with the requirements of direct real-time target tracking. This has been addressed by recent advances in the field of real-time MR-image registration allowing to process a continuous 2D MRI-data stream on a voxel-by-voxel basis with a temporal resolution of up to 20Hz and processing latencies of less than 70ms (Roujol et al 2009, Roujol et al 2010).

Another limitation for direct MRI based 3D target tracking is the required temporal resolution, which prevents the acquisition of extended 3D volumes with sufficient spatial resolution. In practice this leads to MR-imaging of either 2D or severely under-sampled 3D data-sets (2-3 slices with modest spatial resolution in the slice direction). For abdominal motion this can be alleviated by aligning the normal vector of the slice orthogonal this motion vector and thus to contain the entire motion cycle within a 2D imaging slice as suggested by (Denis de Senneville et al 2004). This, however, imposes severe constraints on the imaging geometry which might be unfavorable for anatomical or diagnostic reasons. Furthermore, although the motion trajectory of the kidney and the lower part of the liver can be approximated in first order by a linear shift, the true trajectory is a curve in 3D space. In particular the upper part of the liver, which is subject to an elastic deformation, is hard to contain in a static 2D imaging slice during the entire respiratory cycle. Alternatively, recent studies have demonstrated the possibility to combine 2D image based tracking with an independently placed 1D pencil-beam MR-navigator to achieve full 3D motion compensation on real-time acquired MR-data (Ries et al 2010).

The main disadvantage of direct motion tracking is the requirement to adapt the MRI acquisition with respect to slice positioning, temporal and spatial resolution to the observed motion pattern. Since for most application scenarios MRI serves a dual purpose, motion tracking and MR-thermometry, this potentially limits the performance of MR-thermometry, where a higher spatial resolution and volume coverage is generally preferable to a very high temporal resolution.

Alternatively, direct target tracking could, in principle, also be achieved using real-time Ultrasound-Imaging instead of MRI: Real-time target tracking based on ultrasound echoes with a low processing latency has been successfully demonstrated by Pernot et al (Pernot et al 2004). This study used 4 sonar receivers to estimate the 3D displacement of the targeted organ (3 transducers were required to estimate the displacement, and another one was added to increase the robustness of the process) to achieve dynamic beam steering of a HIFU-system on a moving target. Also the combination of real-time MRI with real-time US-imaging has been successfully demonstrated by several groups: Günther et al (Günther and Feinberg et al 2004, Feinberg et al 2010) combined MRI with simultaneous US-imaging for dynamic MRI-slice tracking. Recent studies (Lourenço de Oliveira et al 2010) demonstrated successfully the combination of 1D motion tracking based on ultrasound echoes for HIFU beam steering on a moving target in combination with simultaneous real-time MR thermometry.

Although the combination of both imaging modalities shows great potential for real-time beam steering and MR-thermometry, it has currently not yet been demonstrated to be feasible in a clinical application scenario. In the presented studies, the motion estimation with ultrasound echoes has been restricted to translational motion and not to observe the ablation area directly, due to the echo perturbation induced by the temperature rise. In addition, signal interference with the therapeutic HIFU system and the partial obstruction of the target area due to ribs and/or air in the beam path has to be addressed in the future.

3. **MR-thermometry and dosimetry of moving organs**

The principal role of MR-thermometry for the guidance of HIFU ablations on liver and kidney is the continuous monitoring of the ablation progress of the intervention for increased patient safety. In addition, MR-thermometry can also be used to provide necrosis estimates and thus to determine the therapy endpoint. Furthermore, thermometric information can be used for adaptive ablation strategies, which employ feedback control of the HIFU power and dynamic modifications of the HIFU trajectory.

Although several methods have been proposed in the past for MR temperature mapping (see the non-exhaustive list of reviews (McDannold et al 2000), (Quesson et al 2000), (Moonen et al 2001), (Jolesz et al 2002), (Salomir et al 2005), (Kuroda et al 2005), (McDannold et al 2005), (Stafford and Hazle 2006), (Denis de Senneville et al 2007), (Rieke and Pauly 2008) for a comprehensive overview), the most frequently used approach for this role is water proton resonance frequency shift (PRF) MR-thermometry. PRF MR-thermometry is based on the linear dependence of the water proton resonance frequency on the local temperature (Ishihara et al 1995). However, the water proton resonance frequency depends also on several other factors, such as the local magnetic field and the local magnetic susceptibility, which even in a well shimmed MR-system varies considerably across the abdomen. As a consequence, PRF MR-thermometry on abdominal organs requires in practice a calibration of these spatial variations in order to achieve acceptable accuracy.

3.1. Correction strategies for spatially inhomogeneous magnetic fields

### 3.1.1. Internal reference

Due to the high fat content of liver tissue and the independence of the proton resonance frequency of hydrocarbons on the temperature, the fat signal has been suggested as a means to provide a calibration of the local magnetic field (Kuroda et al 1996), (Kuroda 2005), (Sprinkhuizen et al 2010). This requires the simultaneous acquisition of the water proton-resonance signal and the hydrocarbon proton resonance frequency, whereby the former serves as a probe for the local temperature and the latter as a temperature independent probe for the local magnetic field. The advantage of this approach is that the magnetic field is measured simultaneously and at the same spatial location as the signal serving as a thermometric probe. This leads to a high degree of accuracy and the possibility to obtain non-invasively absolute temperature measurements. The disadvantage of the approach is the requirement to acquire in addition to the three spatial dimensions an additional spectral dimension. This spectral dimension has to provide sufficient resolution and bandwidth, to permit both the separation of the two signals and to cover the dynamic ranges of the temperature induced frequency changes (Sprinkhuizen et al 2010). Compared to alternative approaches based on 3D imaging, this leads to prolonged acquisition times, which compromises the achievable spatio-temporal resolution and thus renders the approach in practice currently challenging for interventional guidance on moving organs.

### 3.1.2. Preparational calibration scan as reference

The term "Referenced PRF- MR-thermometry" is used in the literature for PRF MR-thermometry which obtains a magnetic field map prior to the heating process as a reference. Under the assumption that only temperature related changes to PRF occur during the subsequent intervention, the local frequency difference is directly proportional to the relative temperature change between both measurements. In practice this allows to map the local water proton resonance frequency using phase images obtained with a gradient recalled MR-sequence at a fixed echo time, and to calculate the relative temperature change by calculating the scaled difference of the phase images (De Poorter et al 1995). The majority of the published *in vivo* studies achieved this by using spoiled gradient recalled MR-sequences in combination with an echo-planar readout to optimize the sampling efficiency, since an optimal thermometric precision requires echo-times close to the $T_2^*$ relaxation time of the target area (Chung et al 1996). Contrary to internally referenced PRF MR-thermometry, this approach requires the signal from other metabolites than water to be discarded for accurate measurements (Kuroda et al 1997, de Zwart et al 1999) and allows only monitoring relative temperature changes from the temperature distribution in the reference.

The main drawback of this approach is that all magnetic field changes between the reference and thermometric measurement, for example due to motion (see section 3.2.2) or instrument drifts (El-Sharkawy et al 2006), are incorrectly attributed to a shift of the water proton resonance frequency due to temperature (Peters et al 1998).

### 3.1.3. Magnetic field modeling

"Referenceless" PRF MR-thermometry corrects the variations of water proton resonance frequency due to local magnetic field inhomogeneities by a direct estimation of the local magnetic field. Contrary to "Referenced" PRF- MR-thermometry, which achieves this by acquiring an initial field map, this approach estimates the inhomogeneous magnetic field in the heated target area from frequency measurements from adjacent unheated tissue. The initial implementation suggested by Vigen et al achieved this using a polynomial extrapolation of the image phase of the surrounding tissue across the target area (Rieke et al 2004). The approach is based on the assumption that the local susceptibility distribution in large organs such as the liver is homogeneous, and consequently the magnetic field in a well shimmed MR-system varies spatially only slowly across the organ. More complex approaches estimate the magnetic field in the target area by numerically solving the $2^{nd}$ Maxwell equation (Salomir et al 2003). The required boundary condition is provided by a field measurement in the surrounding isotherm area at baseline temperature. The main advantage of this approach is the possibility to regularly update the local magnetic field estimation and thus to compensate for magnetic field fluctuations arising from motion of instrument drifts. Limitations of the approach arise from the requirement of a perimeter in which artifact free phase measurements at baseline temperature can be obtained, which is generally problematic in the vicinity of organ boundaries or smaller structures (Denis de Senneville et al 2010).

### 3.2. Influence of motion on PRF MR-thermometry

Unfortunately, PRF MR-thermometry of moving targets, such as the liver and the kidney, is complicated by the continuous target motion through an inhomogeneous and time-variant magnetic field. This effect induces an apparent temperature modification which can bias or even completely mask the true temperature evolution induced by the energy deposition (Peters and Henkelman 2000). Several correction strategies have been proposed in the past, such as respiratory gating (Moricawa et al 2002), navigator echoes (de Zwart et al 2001), multi-baseline acquisition to sample periodic changes (Vigen et al 2003, Denis de Senneville et al 2007, Hey et al 2009) and referenceless phase corrections (Rieke et al 2004).

Furthermore, although the local temperature is a precise indicator for the local energy deposition, it does not directly allow to estimate tissue damage and thus to determine the therapy endpoint. For this purpose, the concept of the equivalent thermal dose was introduced to reflect the accumulative biological effects of sustained elevated temperatures on tissue by (Sapareto and Dewey 1984). In this model the tissue destruction is achieved when the equivalent thermal dose exceeds the lethal dose (which is taken as 43 °C during 240 mn).

Since the calculation of the thermal dose requires a temporal integration of the temperature on a voxel-by-voxel basis, the effect of organ motion has to be addressed either by suitable respiratory gating strategies, or by applying a precise image registration on the temperature maps in order to assure a common voxel position of each part of the observed anatomy (Denis de Senneville et al 2010).

### 3.2.1. Respiratory gated MR-thermometry on kidney and liver

Respiratory gated acquisition strategies for MR-thermometry on kidney and liver have been successfully employed by several groups in past (Vigen et al 2003), (Weidensteiner et al 2004), (Okada et al 2006), (Rempp et al 2011) as shown in Figure 3.

The respiratory gating omits the requirement of complex motion and phase correction strategies and thus facilitates a simple and robust implementation of referenced PRF-Thermometry. Furthermore, the exhalation phase in the respiratory cycle allows under free-breathing conditions an acquisition window of 1000-2000ms. In conjunction with state-of-the-art MR-thermometry sequences this allows to achieve an excellent precision, volume coverage and spatial resolution (Rempp et al 2011).

When applied over extended durations of several minutes, several modifications have been suggested to improve the accuracy of the thermometric measurements: Changes of the respiratory induced motion pattern and sensor drifts, which in turn can modify the gating point and long-term drifts of the local magnetic field, have been addressed by additional phase corrections (Seror et al 2007). The removal of residual spatial misalignment using additional on-line image registration has been suggested (Vigen et al 2003), but so far not been demonstrated in clinical studies.

The main disadvantage of this approach is that the temporal resolution is locked to a multiple of the frequency of the respiratory cycle. It has been demonstrated that gated MR-thermometry is sufficient for the guidance of HIFU generators with moderate power output (<200W), which are limited in their maximum induced temperature increase in the target tissue to less than 1-2°C per second (Lepetit-Coiffé et al 2010). However, the increasing consensus is to perform HIFU ablations in highly perfused organs such as the liver and the kidney with high-power HIFU systems (>350W), allowing a much more aggressive heat-up phase with a temporal temperature evolution of 10-15°C per second. For such application scenarios, gated MR-thermometry is less suitable since it shows the tendency to temporally under-sample the temperature evolution. This is particularly limiting if MR-thermometry is employed for retroactive feedback control strategies of the HIFU power and ablation trajectory.

### 3.2.2. Continuous MR-thermometry of kidney and liver

Continuous, or non-gated, MR-thermometry circumvents the limitations of gated approaches with respect to the temporal resolution by continuous sampling during the respiratory cycle. This allows choosing the optimal compromise between temporal resolution, the spatial resolution, the volume coverage and the thermometric precision for each particular application scenario. While a temporal resolution of 1-2Hz has been reported as sufficient for most applications (Quesson et al 2011), it has been demonstrated that for time-critical applications, continuous real-time MR-thermometry is able to achieve a temporal resolution of 10-15Hz while maintaining a spatial resolution of 2.5x2.5x5mm and a thermometric precision of ±2ºC (Roujol et al 2010). However, the flexibility and performance with respect to the temporal resolution of continuous MR-thermometry on kidney and liver introduces several technical challenges and limitations:

#### 3.2.2.1. Intra-scan artifacts

Real-time MRI of liver and kidney is in general limited by the restricted acquisition time due to the displacement of the organ and the limited signal level due to the short $T_2^*$ relaxation time of the liver. The exploitable acquisition time for a Fourier encoded imaging slice is limited to 75-100ms under free-breathing conditions during ex- or inhalation. Longer periods require complex correction strategies to prevent intra-scan artifacts. As a consequence, most studies employed rapid spoiled gradient recalled MR-sequences with an additional echo-planar readout for optimal the sampling efficiency (Rieke et al 2004 , Roujol et al 2010, Quesson et al 2011 ).
Similar to conventional diagnostic real-time MRI on moving organs, several methods for accelerated MRI have been suggested in addition in order to increase volume coverage and spatial resolution, while remaining within the exploitable sampling window: Parallel imaging such as SENSE (Pruessmann et al 1999) has been used for this role (Bankson et al 2005, Delabrousse et al 2010), but is compared to diagnostic MRI currently limited by a lack of optimized parallel imaging receiver arrays, which are HIFU compatible (i.e. the coil must not obstruct the acoustic propagation path, which considerably complicates the design. In addition, advanced kt-sampling methods have been successfully demonstrated to achieve an increased spatio-temporal resolution for continuous MR-thermometry such as UNFOLD (Mei et al 2011) or a temporally constrained reconstruction (TCR) (Todd et al 2009).

#### 3.2.2.2. Inter-scan artifacts

Since real-time MR-guidance of HIFU ablations requires temperature information predominantly in the vicinity of the ablation area, almost all the currently proposed MR-acquisition schemes sacrifice spatial coverage for an increased spatio-temporal resolution and thermometric precision. As a consequence, most published studies acquired between 1 and 5 slices with modest spatial resolution of 5-7mm in the slice direction with each dynamic scan. This in turn introduces the risk of through plane motion during the respiratory cycle. Most published studies have addressed this problem by careful slice alignment: Although kidney and liver move during the respiratory cycle along a 3D trajectory, the dominating motion component is a linear shift of 1-2cm in head-foot direction, while the remaining components are of much smaller magnitude. As a result, most studies have addressed the problem of through-plane motion successfully by careful slice alignment with the principal displacement vector. Alternatively, the combination of fast in-plane image realignment in conjunction with dynamic slice tracking using pencil-beam navigator echoes has been demonstrated to be able to achieve a full 3D correction (Ries et al 2010, Köhler et al 2011).

Another inter-scan artifact introduced by the continuous sampling during the respiratory cycle is in-plane motion. For applications which employ MR-thermometry only for the visual control of the progress of the intervention, the removal of in-plane motion in real-time is not strictly necessary. Applications which use the temperature maps as the basis for retroactive feedback control of either the HIFU power and/or the HIFU trajectory, generally need to evaluate the spatio-temporal evolution of the temperature in the target area on a voxel-by-voxel basis. Similarly, the calculation of the thermal-dose used for an on-line necrosis prediction requires the temporal integration of the temperature on a voxel-by-voxel basis (Denis de Senneville et al 2007). As a consequence, an in-plane motion compensation in real-time becomes for the later applications a necessary prerequisite.

The most severe inter-scan artefact for non-gated MR-thermometry is caused by the fact that the phase images, which are required for the PRF-method, are acquired at different positions during the passage of the organ through a spatially inhomogeneous and time-variant magnetic field. The resulting thermometric artifacts have typically (Rieke and Butts Pauly 2008) a magnitude which completely masks the true underlying temperature change. As a consequence, this effect requires extending the magnetic field correction strategies described in section 3.1 from only spatially variant fields to spatial-temporal variant fields.

### 3.2.2.2.1. Multi-baseline calibration as magnetic field reference

The multi-baseline correction is a direct extension of the "referenced" PRF- MR-thermometry, which is applied on static objects. The approach exploits the inherent periodicity of the respiratory cycle: A calibration scan performed prior to the intervention is used to establish a lookup-table of magnetic field maps. For that purpose, the motion cycle must be sampled with a sufficient density in order to avoid discretization errors. The field maps are tabled in conjunction with a unique identifier, which allows associating every acquired field map with the corresponding respiratory state. Several possibilities have been proposed for this unique identifier: Navigator echoes (Vigen et al 2003), anatomical MR-images (Denis de Senneville et al 2004), pencil-beam navigator echoes (Hey et al 2009). During the subsequent thermal therapy, this unique identifier allows to recover the field correction which is associated with the current respiratory state. Similar to indirect tracking approaches described in chapter 2.2.1, the main advantage of this approach is that computationally intensive processing steps are separated from the therapeutic part of the intervention, which is conducted under MR-guidance. This allows to achieve very high temporal resolution with very short processing latencies, which are beneficial for the stability of closed loop retroactive control loops for power and trajectory control of the HIFU system: Roujol et al (Roujol et al 2010) demonstrated to be able to provide temperature maps with precision of ±2°C while maintaining a spatio-temporal resolution of 2x2x5mm at 10Hz.

The principal limitation of this approach is that thermometric artifacts associated with spontaneous motion can intrinsically not be corrected. The observation of a significant deviation from the pre-recorded motion pattern requires discarding the image data and, for the case that the change is not reversible, leads to the necessity to interrupt the intervention and to recalibrate the phase correction data (Denis de Senneville et al 2010). Another drawback of Multi-baseline MR-thermometry compared to the referenceless methods is that long-term magnetic field drifts (El-Sharkawy et al 2006) are not inherently corrected, which generally has to be addressed by an additional drift correction.

Several extensions have been proposed to improve the original multi-baseline approach: Hey et al (Hey et al 2009) proposed a linear interpolation avoid discretisation artefacts associated with calibration data of too sparse sampling density of the respiratory cycle. Roujol et al (Roujol et al 2010) combined the correction with in-plane motion stabilization to allow thermal dose calculations and facilitate retroactive feedback control. Denis de Senneville et al (Denis de Senneville et al 2011) used a principal component analysis (PCA) to obtain a parametric spatio-temporal model of the field corrections, which improves the possibility to correct the field for positions which have not been pre-recorded in during the field calibration.

### 3.2.2.2.2. Referenceless approach

The referenceless approach can be designed to intrinsically cope with motion related errors on thermal maps. For that purpose, a dynamic recalculation of the magnetic field is required for each new temperature image (Rieke et al 2004). For non-invasive ablations deep within abdominal organs, the referenceless approach was found to be a robust and fast solution (Holbrook et al 2009), which is not hampered by time-consuming preparations and the complexity of additional corrections for the effects of temporal drifts or occasional spontaneous motion. Its main disadvantage, the requirement of a perimeter in which artifact free phase measurements at baseline temperature, limit it's applicability in areas with strong local susceptibility variations and on organ boundaries.

### 3.2.2.2.3. Hybrid approaches

Due to the fact that the main limitations of multi-baseline and referenceless MR-thermometry are largely complementary, several hybrid approaches have been proposed to overcome these limitations. A direct combination of both techniques to compute each temperature map has been proposed (Grissom et al 2010), as well as a temporal switch method (Denis de Senneville et al 2010). The later employs initially the multi-baseline algorithm to continuously provide temperature maps across the entire field of view. Should a spontaneous movement occur during the intervention, for which no reference phase is pre-recorded, the processing pipeline switches dynamically from multi-baseline to referenceless MR-Thermometry.

### 4. Heterogeneous tissue / Intracostal firing

The non-invasive treatment of liver and kidney tumors with HIFU is complicated by the fact that the acoustic beam path traverses tissues with different celerity and acoustic absorption. Celerity differences in the tissue composition of the beam path generally lead to focus aberrations, focus shifts and partial reflection of the incident beam. This can lead to undesired tissue damage (focus shifts and reflection) and inefficient heating (focus aberrations) (Khokhlova et al 2010).

In particular ablations in the superior part of the abdomen, which is partially obstructed by the thoracic cage, are difficult to achieve. Compared to human soft tissues, bone tissue is a strong absorber for ultrasonic waves. As a consequence, an ablation across the thoracic cage is hampered by two additional effects: First, the absorption of the acoustic energy in the thoracic cage leads to undesired heating of the bone marrow and the adjacent tissue. Second, the obstruction of the free propagation of the focused ultrasonic waves leads to energy loss in the focus and degrades the focus quality.

Two approaches are currently under active investigation to improve the focus quality of intercostal sonications, while simultaneously reducing the acoustic illumination of the ribs. However, since the target location and the exact anatomical structure in the beam path vary considerably between patients, both optimization approaches have be conducted inter-individually and for large ablation areas for different target locations separately.

### 4.1. Improved intercostal firing using acoustic simulations

Acoustic simulations based on a precise model of the acoustic propagation path allow in principle to calculate an optimal aperture function for the transducer, which minimizes undesirable energy deposition in the thoracic cage and recovers the maximal possible focus quality. This approach requires as a first step a precise characterization of the heterogeneous anatomical structures present in the beam path. Subsequently, the anatomical structures are associated with the corresponding acoustic properties (absorption, celerity), which allows to simulate the beam propagation for each transducer element.

Due to the complexity of the correction method, initial approaches neglected the influence of celerity differences (i.e. diffraction and refraction effects) and addressed the undesirable energy deposition in the thoracic cage first. Liu et al (Liu et al 2007) employed CT images from patients to characterize the exact 3D structure and location of the thoracic cage with respect to the transducer and the target area. Subsequently, in his simple straight-line ray-tracing approach all transducer array elements whose normal vectors intersect with the thoracic cage are deactivated. The effectiveness of this approach was demonstrated through numerical simulations, which reconfirmed a reduction of the temperature elevation at the thoracic cage. Quesson et al (Quesson et al 2010) suggested to use anatomical MR images to characterize the structure and location of the thoracic cage and proposed a similar technique in which the shadow of the ribs with respect to the focal point is projected onto the transducer surface by ray tracing from the focal point (Figure 4). Using MR-thermometry, they demonstrated the feasibility to maintain the acoustic energy deposition at the focal point upon deactivation of the shadowed transducer elements. Both approaches are designed to have limited complexity in order to be potentially feasible in the scope of a routine interventional pre-planning. As a consequence, neither wave diffraction by the ribs, nor shear mode conversion have been taken into account.

This might be addressed in the future by more sophisticated acoustic simulations: Botros et al performed simulations to determine the excitation function of the HIFU transducer (Botros et al 1997, Botros et al 1998), Civale (Civale et al 2006) used a linearly segmented transducer and reported the results from acoustic field simulations and measurements on *ex vivo* ribs. More recently, Khokhlova et al demonstrated that for characteristic intensity outputs of modern HIFU arrays, nonlinear effects play an important role and shock fronts develop in the pressure waveforms at the focus (Khokhlova et al 2010, Yuldashev and Khokhlova 2011). A numerical model, based on the solution to the Westervelt equation, was proposed to simulate three-dimensional nonlinear fields generated by high intensity focused ultrasound (HIFU) arrays. The developed algorithm makes it possible to model nonlinear pressure fields of periodic waves in the presence of shock fronts localized near the focus. However, sophisticated acoustic simulations are in general numerically intensive. Since for clinical applications the entire optimization process has to be conducted on an inter-individual basis in the time-frame of the routine interventional pre-planning (i.e. several minutes), much effort has still to be invested in performance optimization, to achieve the correction within a reasonable computation time.

### 4.2. Improved intercostal firing based on direct acoustic measurements

An alternative approach based on direct acoustic measurements is the time-reversal focusing (Tanter et al 2007), based on the concept of the time-reversal mirror introduced by Thomas and Fink

(Thomas et al 1996). Similar to correction based on acoustic simulations, this approach compensates the phase aberration induced by ultrasound propagation through a heterogeneous medium in order to retrieve the optimal acoustic intensity at the focal point. However, the method relies on an initial experiment in which the ultrasound wave front emitted from the focal point by an inserted probe is measured by the HIFU transducer elements. Although the initial approach required invasive measurements, a refined adaptive optimization method (Cochard et al 2009) based on the analysis of the backscattered signals towards the HIFU transducer has been proposed, which does not require direct acoustic measurements at the position of the focal point.

## 5. Volumetric ablations and retro-active feedback control

In general, the tumor volume exceeds the size of the focal point of clinical HIFU systems and is embedded in a highly perfused tissue with a high heat evacuation rate. Since a complete destruction of the tumor is required to assure therapeutic success, efficient ablation control strategies are required exploiting both electronic beam steering and mechanical displacements of the transducer.

The most intuitive approach, a point-by-point ablation, leads to a long overall duration of the intervention due to the required treatment overlap of the ablation points. As a result, volumetric sonications using either mechanical or electronic beam steering along a three-dimensional trajectory have been proposed (Mougenot et al 2008, Köhler et al 2009, Quesson et al 2011), which significantly increase the ablated volume per unit of applied applied acoustic energy (see Figure 5). However, the size of the ablated volume per sonication remains limited by the temperature increase induced by near-field heating in the acoustic beam path in the hypodermis (Damianou and Hynynen 1993, Mougenot 2011). It is currently addressed by introducing cooling intervals between sonications resulting in an increase of the treatment time.

In order to take into account dynamic changes in the energy deposition and heat evacuation during the ablation (acoustic absorption, heat diffusion, tissue perfusion), the concept of volumetric sonications has been combined with retroactive feedback control of the HIFU power (Mougenot et al 2004) and/or the form of the ablation trajectory (Mougenot et al 2008, Enholm et al 2009) (see Figure 6). However, the application of retroactive feedback controlled volumetric sonication strategies in moving targets such as the kidney or the liver, require the approach to be combined with real-time 3D motion compensation for both MR-thermometry and the HIFU beam, which has been demonstrated by Ries et al (Ries et al 2010) in a pre-clinical study.

## 6. Conclusion

MR guided HIFU of mobile organs as kidney and liver is challenging, primarily because of 1) the complex motion and tissue deformation during the respiratory cycle leading to artifacts in MR temperature mapping and requiring real-time HIFU target tracking, 2) the presence of ribs partially blocking the HIFU beam from transducer to target, and 3) the high perfusion rate leading to rapid heat evacuation. Recent technological improvements, described above, have led to significant improvements in MR temperature mapping of these organs, in motion compensation of the HIFU beam, in intercostal HIFU sonication, and in volumetric ablation and feedback control strategies. Recent pre-clinical studies have demonstrated the feasibility of each of these novel methods individually. The perspectives to translate those advances into the clinic look promising but require integration of the different modules, and acceleration of image processing. It can be concluded that MR guided HIFU for ablation in liver and kidney is feasible but requires further work on the combination of several technologically advanced MRI and HIFU methods with feedback coupling.

## Outlook

**Abbreviations:**

HIFU: High Intensity Focused Ultrasound
PRF: Proton Resonance Frequency
TCR: Temporally Constrained Reconstruction
SENSE: Sensitivity Encoding
UNFOLD: Unaliasing by Fourier-encoding the overlaps in the temporal dimension

**Figures**

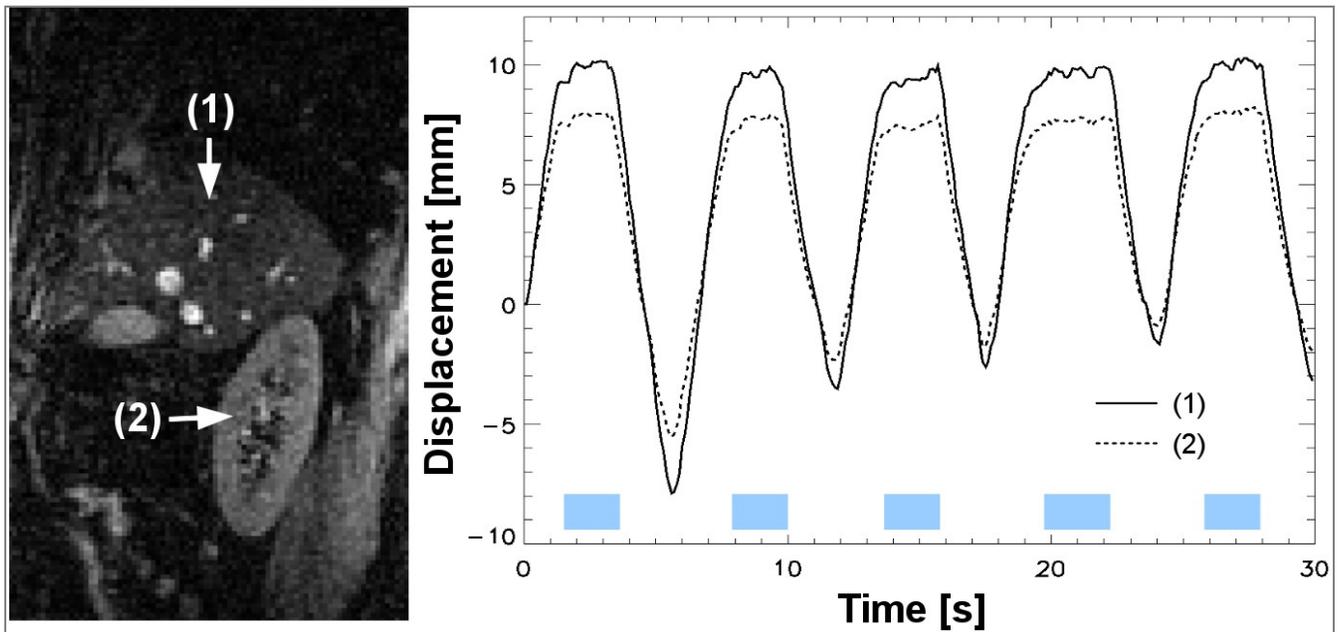

**Figure1.** Comparison of respiratory gated and continuous motion compensated sonication strategies on abdominal organs. The estimation of the organ motion is performed using the sagital anatomical images (reported on the left) using real-time image processing as described in (Denis de Senneville et al 2011). The vertical component of the estimated motion is reported on the right in the liver (1) and the kidney (2). The HIFU sonication duration with the gated sonication strategie (~2 s) is represented by the blue rectangles and cover ~ 1/3 of the total experiment time.

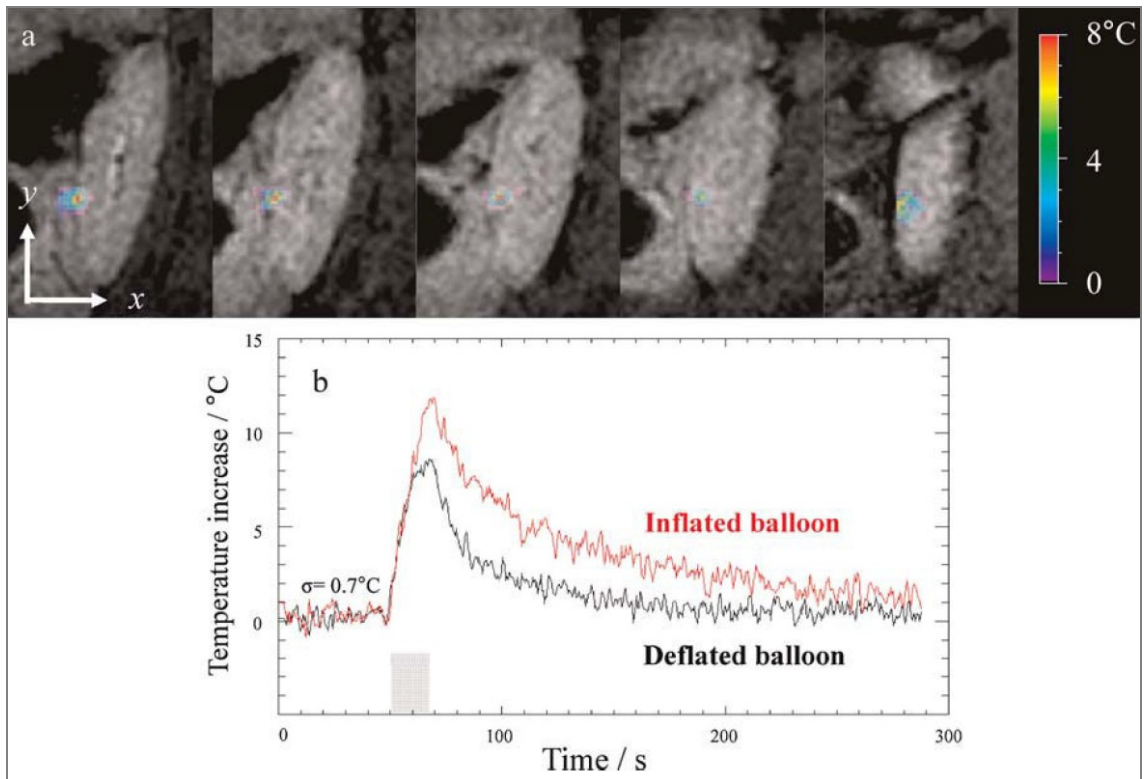

**Figure 2.** Quantitative *in vivo* evaluation of the tissue thermal properties during high-intensity focused ultrasound (HIFU) heating. For this purpose, a total of 52 localized sonications were performed in the kidneys of six pigs with HIFU monitored in real time by volumetric MR thermometry. The kidney perfusion was modified by modulation of the flow in the aorta by insertion of an inflatable angioplasty balloon. The resulting temperature data were analyzed using the bio-heat transfer model in order to validate the model under *in vivo* conditions and to estimate quantitatively the absorption, thermal diffusivity and perfusion (wb) of renal tissue. An excellent correspondence was observed between the bio-heat transfer model and the experimental data. The absorption and thermal diffusivity were independent of the flow, with mean values (± standard deviation) of 20.7 ± 5.1 mm3 K J-1 and 0.23 ± 0.11 mm2 s-1, respectively, whereas the perfusion decreased significantly by 84 % (p<0.01) with arterial flow (mean values of wb of 0.06 ± 0.02 and 0.008 ± 0.007 mL-1 mL s-1), as predicted by the model. The quantitative analysis of the volumetric temperature distribution during nondestructive HIFU sonication allows the determination of the thermal parameters, and may therefore improve the quality of the planning of noninvasive therapy with MR-guided HIFU. (a) Temperature images superimposed on the magnitude images for a HIFU sonication performed in the renal cortex. The scale for the temperature elevation is indicated on the right. The anterior–posterior and inferior–superior directions are represented by 'x' and 'y' arrows, respectively, on the first image. (b) Temperature evolution at the focal point without (black) and with (red) inflating the balloon. The standard deviation of the temperature (s) reported on the graph was calculated from the 40-s MR thermometry data acquired prior to sonication. The gray rectangle shows the sonication duration.
Reproduced with permission of (Cornelis et al 2010).

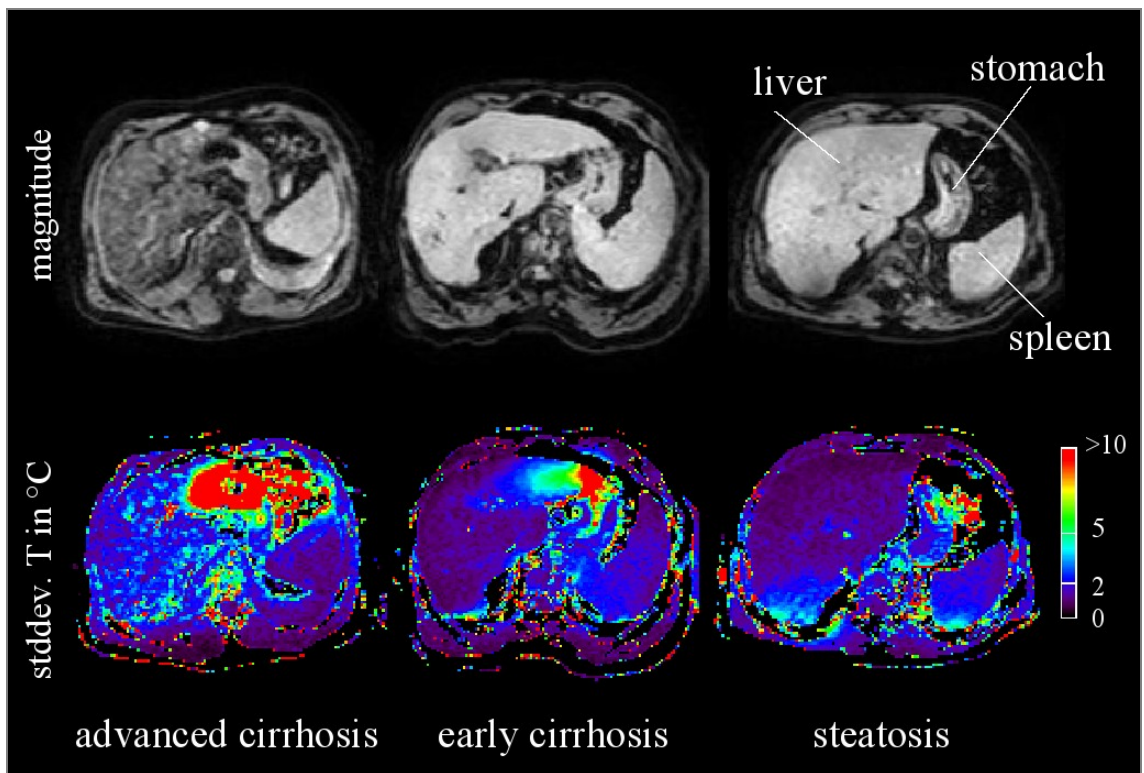

**Figure 3.** Transversal images in 3 liver tumor patients acquired with the respiratory-gated SENSE-EPI sequence, together with corresponding color-coded map of the temperature standard deviation of a time series of temperature maps obtained at normal body temperature.
Reproduced with permission of (Weidensteiner et al 2004).

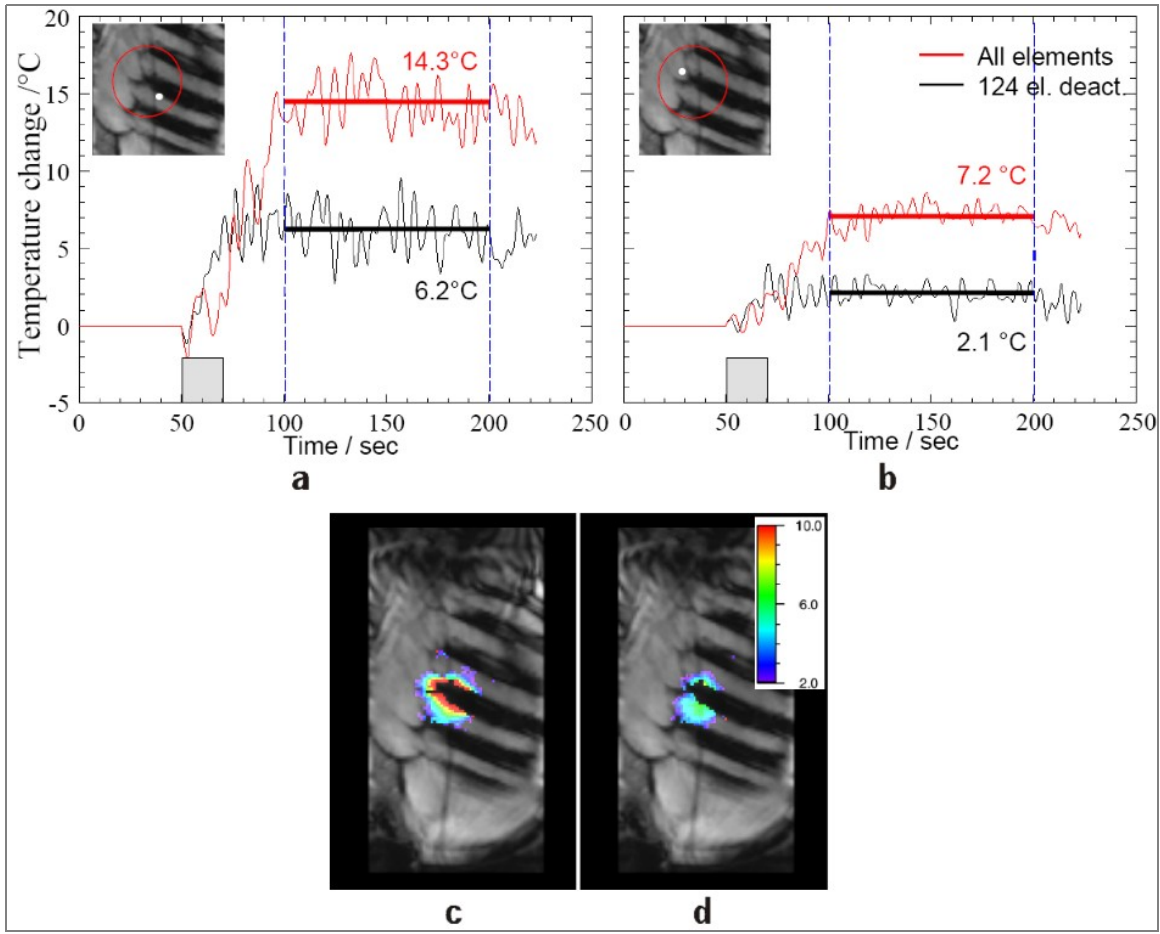

**Figure 4.** Comparison of temperature data near the ribs during HIFU sonications performed in pig liver *in vivo*, without and with deactivation of the transducer elements. The graphs show the temperature evolution in a single pixel located (a) near the rib and (b) in the cartilage (see white points for exact location in each insert). The red curves show the temperature evolution when all the HIFU elements are active and the black curves show the temperature evolution when 124 elements were deactivated. The HIFU sonication duration (20 s) is represented by the gray rectangles. The horizontal bars show the average values of temperature between 100 and 200 seconds. Images (c) and (d) are maps of the average temperature values (the color scale is indicated on the right) in each pixel between 100 and 200 s (vertical dashed blue lines in a and b).
Reproduced with permission of (Quesson et al 2010).

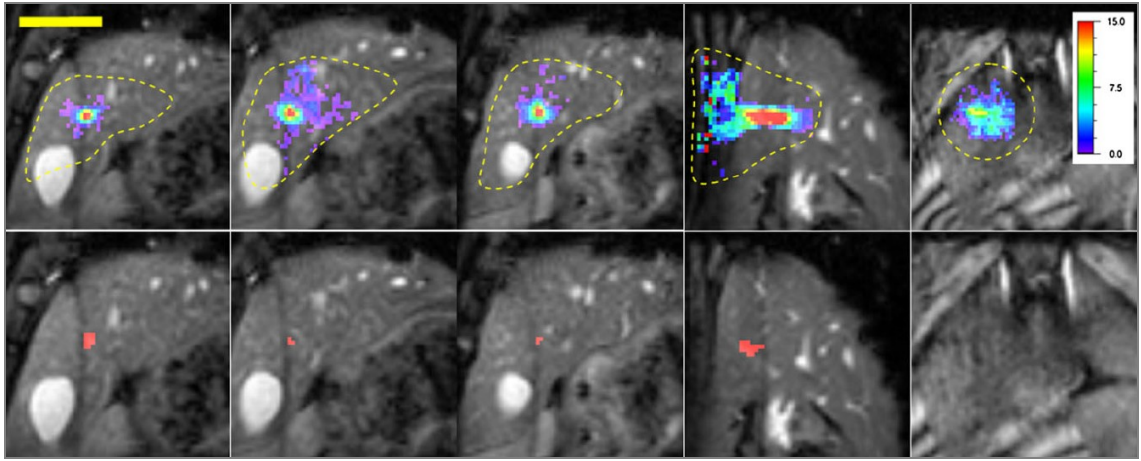

**Figure 5.** Temperature (top row) and thermal dose (bottom row) images of the liver during high-intensity focused ultrasound. The temperature scale is displayed on the right. The pixels colored in red correspond to a thermal dose of 240 TEM43C and above. Left to right: three coronal slices centred around the target point, one sagittal slice centred in the target point and one slice located near the skin. The broken contours in the temperature images show the selected regions of interest for displaying the temperature data. The temperature images display the temperature distribution measured at the end of HIFU sonication, and the thermal dose images are the final images in the time series.
Reproduced with permission of (Quesson et al 2011).